\documentclass[11pt,preprint]{aastex}

\shorttitle{Habitable Planets Around WDs} 
\shortauthors{Kilic et al.}

\begin{document}

\title{Habitable Planets Around White Dwarfs: \\
       an Alternate Mission for the Kepler Spacecraft}

\author{Mukremin Kilic\altaffilmark{1,6},
Eric Agol\altaffilmark{2},
Abraham Loeb\altaffilmark{3},
Dan Maoz\altaffilmark{4},
Jeffrey A. Munn\altaffilmark{5},\\
Alexandros Gianninas\altaffilmark{1},
Paul Canton\altaffilmark{1},
Sara D. Barber\altaffilmark{1}
}
\altaffiltext{1}{Department of Physics and Astronomy, University of Oklahoma, 440 W. Brooks St., Norman, OK, 73019}
\altaffiltext{2}{Department of Astronomy, Box 351580, University of Washington, Seattle, WA 98195}
\altaffiltext{3}{Institute for Theory and Computation, Harvard University, Cambridge, MA 02138}
\altaffiltext{4}{School of Physics and Astronomy, Tel-Aviv University, Tel-Aviv 69978, Israel}
\altaffiltext{5}{US Naval Observatory, P.O. Box 1149, Flagstaff, AZ 86002}
\altaffiltext{6}{kilic@ou.edu}

\begin{abstract}
A large fraction of white dwarfs  (WDs) may host planets in their habitable zones. 
These planets may provide our best chance to detect bio-markers on a transiting exoplanet,
thanks to the diminished contrast ratio between the Earth-sized WD
and its Earth-sized planets. The James Webb Space Telescope is capable of obtaining
the first spectroscopic measurements of such planets, yet there are no known planets
around WDs. Here we propose to 
take advantage of the unique capability of the Kepler spacecraft in the 2-Wheels mode
to perform a transit survey that is capable of identifying the first planets in the habitable
zone ($P =$ 4-30 h) of a WD. We propose to obtain Kepler time-series photometry of
$10^4$ WDs in the Sloan Digital Sky Survey imaging area
to search for planets in the habitable zone. Thanks to the large
field of view of Kepler, for the first time in history, a large number of WDs
can be observed at the same time, which is essential for discovering transits. 
Our proposed survey requires a total of 200 days of observing time, and will find up to
100 planets in the WD habitable zone. This survey will maintain Kepler's spirit of
searching for habitable Earths, but near new hosts. With few-day observations and
minute-cadences per field, it will also open up a completely unexplored discovery space.
In addition to planets, this survey is sensitive to pulsating WDs, as well as
eclipsing short period stellar and substellar companions.
These have important implications for constraining the double WD merger rate and
their contribution to Type Ia supernovae and the gravitational wave foreground. Given the
relatively low number density of our targets, this program can be combined with other
projects that would benefit from high cadence and `many-fields' observations with Kepler,
e.g. a transit survey of a magnitude-limited, complete sample of nearby M dwarfs or
asteroseismology of variable stars (e.g. RR Lyrae) in the same fields. 
\end{abstract}

\section {Background}

Transiting exoplanets provide invaluable information on planetary physics, their formation, and 
evolution. The search for planets in the habitable zone has so far focused on solar-type stars
and M dwarfs. However, transiting planets in the habitable zone around white dwarfs (WDs) may
be common and they may provide our best chance to detect biomarkers on an exoplanet in the near future
\citep{loeb13}. 

\begin{figure}[h]
\epsscale{0.7}
\plotone{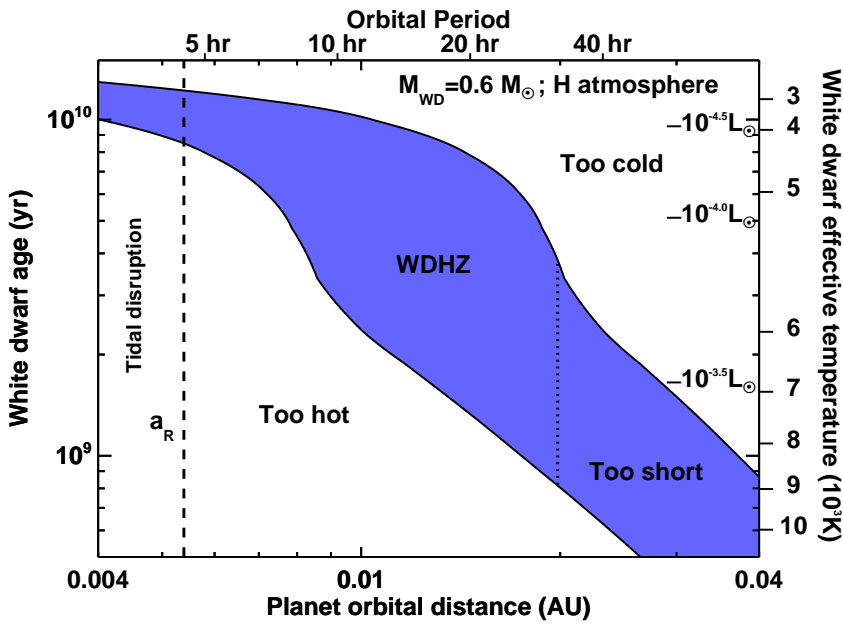}
\caption{The habitable zone (shaded region) around 0.6 $M_{\odot}$ WDs versus age and planet orbital
distance \citep[from][]{agol11}. Dashed line (0.005 AU) is the Roche limit for Earth-density planets.
Planets within 0.02 AU would be in the habitable zone for $>$3 Gyr.}
\label{fig:agol}
\end{figure}

WDs are as common as Sun-like stars, and they provide an energy source for planets for
billions of years. Typical WDs have $M=0.6 M_{\odot}, R=0.01 R_{\odot}$, and $L=10^{-4}L_{\odot}$,
so a planet must orbit at $\sim$0.01 AU to be at a temperature for liquid water to exist on its
surface. The habitable zone around WDs extends from 0.005 AU (P = 4 h) to 0.02 AU
\citep[P = 30 h,][]{agol11,monteiro10} for WDs older than about 1 Gyr. 
Figure \ref{fig:agol} shows the evolution of the habitable zone around WDs (WDHZ) with time.
A planet enters at the bottom of Figure \ref{fig:agol} and moves vertically up the figure as
its WD host ages, so it starts off too hot for liquid water, passes through the WDHZ,
and then becomes too cold. The duration a planet spends within the WDHZ has a maximum of 8 Gyr at
0.01 AU; WDs have long-lived habitable zones.

We expect the planets within 1 AU of solar type stars to be destroyed in the giant phase
\citep[][but see Silvotti et al. 2007]{villaver07}. Hence, planets in the habitable zone around WDs must arrive there after this phase.
\citet{debes02}, \citet{livio05}, \citet{faedi11}, and \citet{veras13}
describe several ways to form such planets near the WD or bring them closer. 
Planets can form out of gas near the WD, via the interaction or merger of binary stars \citep{kratter12},
or by capture or migration from larger distances. 
Planets have been detected around evolved,
post-main-sequence stars, including millisecond pulsars and pulsar + WD systems
\citep{wolszczan92,sigurdsson03,beuermann12}.
\citet{wu11} have a dynamical instability model that can produce short period planets on Gyr timescales; the same mechanism is likely at work after the giant phases and could populate the habitable zone around WDs. 
The discovery of
close-in planets around the post-main-sequence stars KIC 05807616 and V391 Pegasi demonstrate that such planets
should also exist around WDs \citep{charpinet11,silvotti07,passy12}.
\citet{barnes13} and \citet{nordhaus13} have emphasized that the tidal 
heating of a planet, until it achieves full circularization, in the WDHZ would lead to the loss of water.
However, the young Earth was also a hot and dry place, but volatiles and water were then delivered to
it by a large number of comets. It is likely that the process that populates the WD habitable zone could
also lead to a heavy bombardment phase that could deliver water and volatiles to these planets \citep{loeb13}.

About 30\% of the WDs in the solar neighborhood
have metal-polluted atmospheres \citep{zuckerman10}. The source of metals is not accretion from
the interstellar medium \citep{kilic07}. In addition, $\geq$4.3\% of WDs host debris disks 
\citep{barber12}, which are the remnants of tidally destroyed asteroids and planets 
\citep[e.g.,][]{klein11}. Hence, there is direct evidence from these debris disks that the 
interactions between giant planets can 
send asteroids, moons, or small (Earth-like) planets closer to the WD \citep{jura03,veras13}. Therefore, 
short-period planets around WDs likely exist; but we have never looked at enough WDs to find 
them. 

\begin{figure}[h]
\hspace{1.2in}
\includegraphics[width=2.85in,angle=-90]{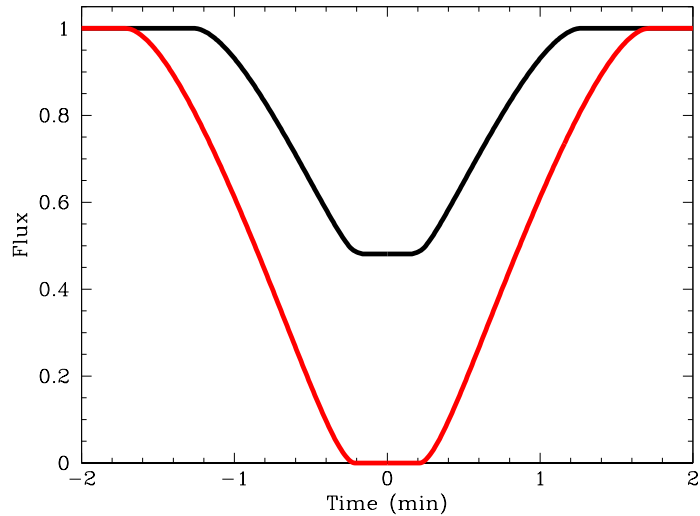}
\caption{Simulation of eclipses caused by a 1 Earth mass planet (top) and a 10 
Earth mass planet (bottom) transiting in the habitable zone (0.01 AU) of a white dwarf.}
\label{fig:eclipse}
\end{figure}

WDs are about the same size as Earth. Hence, earth-size (and even smaller) planets can easily
be detected through photometric observations \citep{distefano10,faedi11,drake10}.
Figure \ref{fig:eclipse} shows model light-curves of a 0.6 $M_{\odot}$ WD 
as it is eclipsed by a planet with $M = 1 M_{\earth}$ and another with 
$M = 10 M_{\earth}$ orbiting in the habitable zone of the WD. 
In the case of the 1 $M_{\earth}$ planet, the eclipse produces a reduction in flux of 
50\%, whereas a 10 $M_{\earth}$ super-Earth would effectively block out the WD
completely. It is this high contrast ratio between the planet and the host WD that makes it
possible to obtain an atmospheric transmission spectrum of planets in the WDHZ with
future telescopes, like the James Webb Space Telescope \citep[JWST,][]{loeb13,agol11}. The most prominent
biomarker on Earth, molecular oxygen (O$_2$), can be detected in the transmission spectrum
of an Earth-like planet orbiting WDs \citep[see Fig. 1 in][]{loeb13}.

To detect these 2-min long transits, one must monitor at high cadence a large sample of WDs
for the duration of
the orbital period at 0.02 AU ($P=30$ h). Assuming random inclinations, the transit probability is roughly
1\% for Earth-size planets. To measure $\eta_{\earth}$, the frequency of Earth-like planets, to an
accuracy of 33\% requires detecting 9 planets, so one must survey $10^3$ WDs. 
Similarly, \citet{loeb13} argue that a survey of at least 500 WDs is required to detect transiting
planets around them. For values of $\eta_{\earth}<$10\%, more stars must be observed 
to detect planets, so the best strategy is to observe multiple WDs simultaneously with
a wide-field imager. Given the unknown frequency of exoplanets in the WDHZ, a survey
of $\sim10^4$ WDs would be essential in constraining the frequency and types of planets in the WDHZ.

A ground-based telescope with a modest field of view (less than a couple sq. deg.) would have to observe the 
stars one at a time. So $10^4$ WDs would require 2-4$\times10^4$ days (including overhead), or
about 100 years. For comparison, Kepler can perform the same survey in 200 days.
There are no planned or ongoing surveys that will accomplish the survey requirements; 30 h long observations
of $\sim10^4$ WDs with 1-2 min cadence. The Transiting Exoplanet Survey Satellite (TESS) is not sensitive
enough to perform a transit survey for a large number of WDs. There are also no ground-based telescopes with sufficient aperture
size and wide field-of-view to perform a survey of $10^4$ WDs and to obtain continuous coverage over 2 days.
For example, the Large Synoptic Survey Telescope (LSST) and
GAIA \citep{perryman01} surveys will obtain 50 to 1000 epochs of observations over a decade, possibly detecting
one epoch in eclipse for
a few per cent of the WDs with transiting planets. These surveys will
be biased toward detecting shorter period and larger planets that have yet to enter the WDHZ.
In addition, confirmation of the planets around many of the faint LSST WDs will be challenging.
On the other hand, the transits detected in our proposed Kepler survey can be easily followed-up
with ground-based telescopes and, most importantly, with the JWST.

\vspace{-0.21in}
\section{The Proposed Survey}

Kepler in the 2-wheels mode provides an unprecedented opportunity to
perform a photometric survey that is capable of finding the first exoplanets 
in the WDHZ. Thanks to its sky coverage and depth, for the first time, Kepler can
image a large number of WDs at the same time and search for transiting planets
and other substellar or stellar companions. The majority of the known WDs are
in the SDSS fields. Hence, Kepler imaging of the known WDs in the SDSS fields
provides the best opportunity for such a transit survey.

\begin{figure}[h]
\hspace{1.2in}
\includegraphics[width=2.85in,angle=-90]{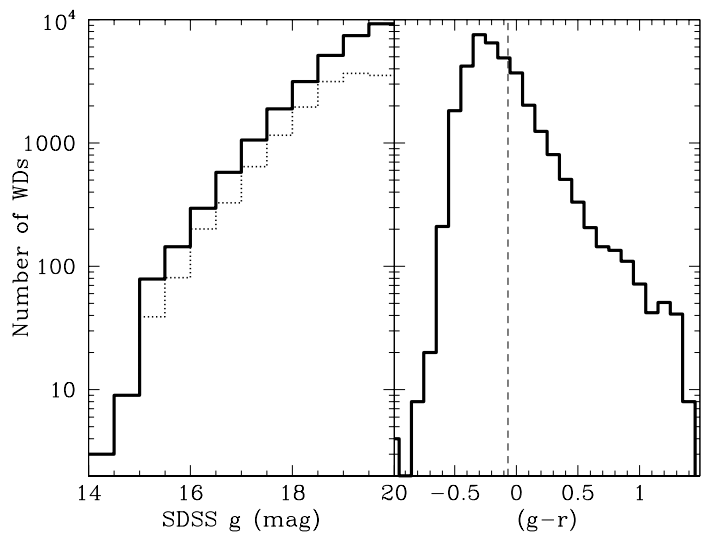}
\caption{Brightness (left) and color (right) distribution of spectroscopically confirmed WDs in the SDSS DR7
\citep[dotted line,][]{kleinman13} plus high proper motion WDs \citep[solid line,][]{munn04}.
There are $10^3$ and $10^4$ targets with $g\leq16.8$ mag and $g\leq18.7$ mag, respectively.
Objects to the right of the dashed-line (right panel) are cool WDs with $T_{\rm eff}\leq$10,000 K.}
\label{fig:sdss}
\end{figure}

There are $\approx2\times10^4$ spectroscopically confirmed WDs in the SDSS DR7 area
\citep{kleinman13}. We have identified an additional $2\times10^4$ WDs
in the same area through proper motion measurements using the SDSS + USNO-B
astrometry \citep{munn04}. Even though many of these targets lack spectroscopy data,
\citet{kilic10} demonstrate that this sample is almost pure, with a
contamination rate from subdwarfs of 1\%. Overall, there are $\approx4\times10^4$
unique WDs currently known in 11,663 sq. deg. of imaging in the SDSS DR7 area.
Figure \ref{fig:sdss} shows the magnitude and color distribution of these targets. 
There are $10^3$ and $10^4$ WDs with $g\leq16.8$ mag and $g\leq18.7$ mag, respectively.
The latter include 2253 cool WDs with $T_{\rm eff}<10,000$ K.
We propose a Kepler imaging survey of the $10^4$ SDSS WDs brighter than $g\leq18.7$ mag
to discover stellar, substellar, and planetary (including sub-earth-and-up companions) around
all WDs. This survey will measure the planetary frequency and survival around both young and old
systems and discover planets in the WDHZ. Thanks to the SDSS astrometry, accurate
positions are known for each target and target apertures can be defined easily.
The limiting factor for Kepler observations of faint sources is set by source confusion, rather
than the photometric accuracy. All of our targets have SDSS imaging down to $g=22$ mag, and the
USNO-B+SDSS proper motion catalog \citep{munn04} avoids sources with nearby background sources
within $7\arcsec$. Hence, source confusion is not a problem for our targets.
The proposed survey will image each field for 2 days, with 1 min cadence. This survey
will be sensitive to planets within 0.03 AU of the stars. Imaging the DR7 area requires
about 100 pointings. Hence, the entire survey can be done over 200 days.

\vspace{-0.16in}
\subsection{Kepler's Sensitivity in the 2-Wheels Mode}

The nominal Kepler mission provided $\sim$50 ppm photometry for a 12 mag G2V star integrated
over 30 minutes, enabling discoveries of Earth-like planets around main-sequence stars. 
Given the pointing problems in the 2-wheels mode, the image will be spread out over more pixels,
and the increased read noise and larger responsivity variations will reduce the photometric precision
to $\sim$0.3-1\%. Given the eclipse signature of Earth-size and larger planets around WDs (see
Figure \ref{fig:eclipse}), the systematic errors due to the pointing problems is not the limiting
factor for WDHZ observations. Instead, the brightness of the target WDs limits the delivered
photometric precision. Kepler can obtain 1\% photometry for targets brighter than about 16.5 mag,
and 10\% photometry for targets brighter than 18.7 mag \footnote{http://keplergo.arc.nasa.gov/CalibrationSN.shtml}. 
Hence, the 50\% eclipse signature of an Earth-like planet can be detected at $\geq5\sigma$ for targets
as faint as 18.7 mag. 
It may be possible to obtain photometry of the fainter targets in the same fields
by using a smaller number of pixels for computing the light curve.
Figure \ref{fig:sdss} shows that the number of objects rises dramatically between
$g=18$ and 20 mag. Hence, the number of targets can easily be doubled by a
$\sim$0.5 mag deeper survey; operational testing of the 2-wheels mode is necessary to decide
on the limiting magnitude for such observations.
Kepler can effectively search for transiting Earth-size planets around $10^4$ WDs in the SDSS.
With its unprecedented photometric sensitivity, Kepler could also detect smaller occulting objects down to the
scale of the Earth's moon, albeit with lower significance.

\vspace{-0.16in}
\section{Added Benefits}

\subsection{Eclipsing WD + M dwarfs: Mass-Radius Relationship}
 
In addition to planetary companions, Kepler will also be sensitive to short period substellar
and stellar companions. This sample will be essentially complete to $P=2$ d. However, eclipsing
systems with longer periods will also be detected. About 22\% of field WDs have late-type
main-sequence star companions \citep{farihi05}, while 3.4\% of these systems are eclipsing
with $P<2$ day orbits \citep{parsons13}. Hence, the frequency of eclipsing WD + M dwarf binaries
is likely around 0.75\%. In a sample of $10^4$ WDs, we are likely
to find 75 eclipsing WD + M dwarf binaries. Earth-density planets will not survive at
orbital periods shorter than 4 h. Hence, eclipses with $P<4$ h are almost certainly from stellar or
brown dwarf companions. 

Given the differences in size, the eclipse signatures of WD + Earth-size
planets and WD + stellar binaries would be significantly different and easy to distinguish based on the
light-curves. In addition, follow-up near-infrared photometry and optical radial velocity
observations of the eclipsing systems can easily distinguish between planetary companions
and brown dwarf or stellar companions. Stellar binaries will show primary and secondary eclipses 
of different depths in different filters, offset secondary eclipses due to light travel time,
gravitational lensing \citep{agol02}, and Doppler beaming \citep{loeb03,zucker07}.
These systems would provide stringent constraints on the mass-radius relations for WDs and
late-type M dwarfs, and also age-activity relation for the latter.

\vspace{-0.16in}
\subsection{Double WDs: SNe Ia Progenitors and Gravitational Wave Sources} 

It is widely accepted that SNe Ia are caused by the thermonuclear explosion
of WDs \citep{webbink84,iben84}, but the nature of
the progenitor binary systems is not settled. Until we find the progenitors,
it is impossible to understand the systematic uncertainties and to optimize
SNe Ia for precision cosmology. The SNe Ia progenitor problem is therefore
a key problem in astronomy today \citep{distefano10,howell11}. Recent theoretical
studies suggest that sub-Chandrasekhar mass WD systems may also form SNe Ia
\citep[e.g.,][]{sim12,pakmor13}. Major contenders for SN Ia progenitors are mergers
of binary WDs. Discovering and quantifying the population properties of
close double WDs, including CO+CO and CO+He binary WD systems, 
can provide key insights into the SNe Ia progenitor problem \citep{badenes12}.

There are currently four eclipsing double WD systems known with orbital
periods ranging from 12.75 min \citep{brown11} to 5.9 h \citep{vennes11}.
The lack of eclipsing systems at longer orbital periods is a selection effect
due to the difficulty of observing a target for longer than 6 h a night.
The proposed Kepler imaging survey will identify all of the eclipsing double WD merger
systems among our targets,
thanks to the 2 day long observations. These eclipses
last 2-3 min, and include total eclipses for similar mass WDs. 
Based on a population synthesis model for binary stars, \citet{agol11} predict
a frequency of 2.5\% for double WDs with $P=8-64$ h.
Our proposed survey will provide the first constraints on the frequency
of eclipsing double WDs and their merger rate. Eclipsing double WDs provide
the most accurate constraints on the masses and radii of each target.
This will in turn provide the mass distribution and merger times for each
system and constrain the contribution of the double degenerate channel to SNe Ia.
This project may finally provide direct evidence for or against the double
degenerate channel for SNe Ia.

The shortest period binary WDs are also excellent gravitational wave sources. The gravitational wave
radiation and the orbital decay in the shortest period systems may
be detected directly by space based missions like the Laser Interferometer Space
Antenna (LISA) and indirectly by ground based observations \citep[see][]{hermes12}.
\citet{nelemans09} lists 12 ultra-compact systems that are guaranteed LISA sources,
but predicts that LISA should detect at least several hundred systems.
Double WDs in the Galaxy outnumber all other known types of gravitational wave sources
and they form a gravitational wave foreground. Some of the eclipsing systems found in the proposed
Kepler survey will be amongst the strongest gravitational wave sources known. With accurate positions,
 masses, and distance estimates, these will be verification sources for gravitational wave
missions in the milli-Hertz range \citep[e.g., eLISA,][]{seoane13}.

\vspace{-0.16in}
\subsection{Pulsating WDs}

WDs go through the DOV, DBV, and DAV instability strips as they evolve. The ZZ Ceti (DAV) stars
are the most common type of pulsators, with $\sim$150 currently known. DAVs
are found in a very well defined region in the $T_{\rm eff}-\log{g}$ plane. They exhibit pulsation
periods in the range 100-1400 s, corresponding to low-degree gravity-mode oscillations. The detected
pulsation modes have amplitudes ranging from a few millimagnitudes to a fraction of a magnitude ($\sim$10\%). 
The cooler pulsators tend to show longer periods and larger amplitudes. Also, for a given effective
temperature, the observed periods tend to be longer for lower gravity objects \citep{gianninas06,hermes13}.
High speed photometric studies of the pulsating WDs enable us to constrain the structure and evolutionary
timescales of WDs. For example, the pulsation modes and amplitudes are sensitive to the thickness of
the surface H and He layers, the depth of the convection zone, and the degree of crystallization \citep[e.g.][]{montgomery10}. 

There are $\sim$1500 WDs with $T_{\rm eff}\approx$12,000 K in our proposed Kepler survey. These WDs
are in the right temperature range to pulsate as a DAV. Kepler will be sensitive to large amplitude pulsations,
and as a bonus, it will provide a magnitude-limited sample of pulsators.  
With 2 d coverage, the
majority of the modes can be identified, and the most interesting targets can be followed up from
the ground for further characterization and theoretical modeling of the structure of each star.
Kepler will significantly increase the number of known pulsating WDs and change this field
dramatically.

\vspace{-0.21in}
\section{Conclusions: Why Should Kepler Observe $10^4$ WDs?}

Before the discovery of hot Jupiters around main-sequence stars, our view of planet
formation was very different. The discovery of hot Jupiters was unexpected and unintentional. 
Here we outlined an alternate mission for Kepler in the 2-wheels mode targeting a large sample
of WDs to find the first planets in the habitable zone. If the history of exoplanet science
has taught us anything, it is that planets are ubiquitous and they exist in the most unusual places,
including very close to their host stars and even around pulsars \citep{wolszczan92}.
Currently there are no known
planets around WDs, but we have never looked at a sufficient number of WDs at high cadence to
find them through transit observations. It is essentially impossible to find Earth-Jupiter size
planets around WDs by any other method \citep{gould08}. If habitable planets exist around WDs, the proposed Kepler
imaging survey will find them. Biomarkers, including O$_2$, on such planets can be detected
with the JWST. Hence, even though this is a completely unexplored search area for transiting planets,
the scientific yield of the proposed survey will be enormous.

In addition to the planetary studies, the proposed survey will also dramatically change the field
of short period binary WDs, including eclipsing double WDs, WD + M dwarfs, and WD + brown dwarfs. 
This survey will constrain the merger rate and mass distribution of double WDs, which is important
for binary population synthesis studies, understanding SNe Ia explosions, and identifying new
verification binaries and the Galactic foreground in gravitational waves.

Short cadence targets are limited to no more than 512 across the entire focal plane. Our proposed
survey has a density of $\sim100$ short cadence targets per pointing. Therefore, this survey
can be combined with other projects targeting the SDSS fields.
For example, similar high (or low) cadence observations of M dwarfs 
\citep[e.g., the Mearth project,][]{berta13} or other types of (variable) stars in the SDSS fields
can be obtained simultenously. Nearby M dwarfs can be easily selected based on their
high proper motion from the SDSS + USNO-B positions \citep{munn04}.
Alternatively, the SDSS photometry can be used to select a nearly complete and contamination free
sample of variable stars like RR Lyrae over the entire SDSS footprint.
The survey can also be extended to an all-sky survey through the selection of WDs from
other proper motion surveys like the USNO-B, SuperCOSMOS \citep{rowell11}, and LSPM \citep{lepine05}.
If this survey is not selected as the primary mission for Kepler in the 2-wheels mode,
it should at least be considered as a filler project for other surveys.



{
\small
\begin{thebibliography}{}
\bibitem[Agol(2002)]{agol02} Agol, E.\ 2002, \apj, 579, 430 
\bibitem[Agol(2011)]{agol11} Agol, E.\ 2011, \apjl, 731, L31 
\bibitem[Badenes \& Maoz(2012)]{badenes12} Badenes, C., \& Maoz, D.\ 2012, \apjl, 749, L11 
\bibitem[Barber et al.(2012)]{barber12} Barber, S.~D., Patterson, A.~J., Kilic, M., et al.\ 2012, \apj, 760, 26 
\bibitem[Barnes \& Heller(2013)]{barnes13} Barnes, R., \& Heller, R.\ 2013, Astrobiology, 13, 279 
\bibitem[Berta et al.(2013)]{berta13} Berta, Z.~K., Irwin, J., \& Charbonneau, D.\ 2013, arXiv:1307.3178 
\bibitem[Beuermann et al.(2012)]{beuermann12} Beuermann, K., Dreizler, S., Hessman, F.~V., \& Deller, J.\ 2012, \aap, 543, A138 
\bibitem[Brown et al.(2011)]{brown11} Brown, W.~R., Kilic, M., Hermes, J.~J., et al.\ 2011, \apjl, 737, L23
\bibitem[Charpinet et al.(2011)]{charpinet11} Charpinet, S., Fontaine, G., Brassard, P., et al.\ 2011, \nat, 480, 496 
\bibitem[Debes \& Sigurdsson(2002)]{debes02} Debes, J.~H., \& Sigurdsson, S.\ 2002, \apj, 572, 556 
\bibitem[Di Stefano et al.(2010)]{distefano10} Di Stefano, R., Howell, S.~B., \& Kawaler, S.~D.\ 2010, \apj, 712, 142 
\bibitem[Drake et al.(2010)]{drake10} Drake, A.~J., Beshore, E., Catelan, M., et al.\ 2010, arXiv:1009.3048 
\bibitem[Faedi et al.(2011)]{faedi11} Faedi, F., West, R.~G., Burleigh, M.~R., Goad, M.~R., \& Hebb, L.\ 2011, \mnras, 410, 899 
\bibitem[Farihi et al.(2005)]{farihi05} Farihi, J., Becklin, E.~E., \& Zuckerman, B.\ 2005, \apjs, 161, 394 
\bibitem[Gianninas et al.(2006)]{gianninas06} Gianninas, A., Bergeron, P., \& Fontaine, G.\ 2006, \aj, 132, 831 
\bibitem[Gould \& Kilic(2008)]{gould08} Gould, A., \& Kilic, M.\ 2008, \apjl, 673, L75 
\bibitem[Hermes et al.(2012)]{hermes12} Hermes, J.~J., Kilic, M., Brown, W.~R., et al.\ 2012, \apjl, 757, L21 
\bibitem[Hermes et al.(2013)]{hermes13} Hermes, J.~J., Montgomery, M.~H., Winget, D.~E., et al.\ 2013, \apj, 765, 102 
\bibitem[Howell(2011)]{howell11} Howell, D.~A.\ 2011, Nature Communications, 2, 350
\bibitem[Iben \& Tutukov(1984)]{iben84} Iben, I., Jr., \& Tutukov, A.~V.\ 1984, \apjs, 54, 335
\bibitem[Jura(2003)]{jura03} Jura, M.\ 2003, \apjl, 584, L91 
\bibitem[Kilic \& Redfield(2007)]{kilic07} Kilic, M., \& Redfield, S.\ 2007, \apj, 660, 641 
\bibitem[Kilic et al.(2010)]{kilic10} Kilic, M., Leggett, S.~K., Tremblay, P.-E., et al.\ 2010, \apjs, 190, 77 
\bibitem[Klein et al.(2011)]{klein11} Klein, B., Jura, M., Koester, D., \& Zuckerman, B.\ 2011, \apj, 741, 64 
\bibitem[Kleinman et al.(2013)]{kleinman13} Kleinman, S.~J., Kepler, S.~O., Koester, D., et al.\ 2013, \apjs, 204, 5 
\bibitem[Kratter \& Perets(2012)]{kratter12} Kratter, K.~M., \& Perets, H.~B.\ 2012, \apj, 753, 91 
\bibitem[L{\'e}pine \& Shara(2005)]{lepine05} L{\'e}pine, S., \& Shara, M.~M.\ 2005, \aj, 129, 1483 
\bibitem[Livio et al.(2005)]{livio05} Livio, M., Pringle, J.~E., \& Wood, K.\ 2005, \apjl, 632, L37 
\bibitem[Loeb \& Gaudi(2003)]{loeb03} Loeb, A., \& Gaudi, B.~S.\ 2003, \apjl, 588, L117 
\bibitem[Loeb \& Maoz(2013)]{loeb13} Loeb, A., \& Maoz, D.\ 2013, \mnras, 432, L11 
\bibitem[Monteiro(2010)]{monteiro10} Monteiro, H.\ 2010, Bulletin of the Astronomical Society of Brazil, 29, 22 
\bibitem[Montgomery et al.(2010)]{montgomery10} Montgomery, M.~H., Provencal, J.~L., Kanaan, A., et al.\ 2010, \apj, 716, 84 
\bibitem[Munn et al.(2004)]{munn04} Munn, J.~A., Monet, D.~G., Levine, S.~E., et al.\ 2004, \aj, 127, 3034 
\bibitem[Nelemans(2009)]{nelemans09} Nelemans, G.\ 2009, Classical and Quantum Gravity, 26, 094030
\bibitem[Nordhaus \& Spiegel(2013)]{nordhaus13} Nordhaus, J., \& Spiegel, D.~S.\ 2013, \mnras, 432, 500 
\bibitem[Pakmor et al.(2013)]{pakmor13} Pakmor, R., Kromer, M., Taubenberger, S., \& Springel, V.\ 2013, \apjl, 770, L8 
\bibitem[Parsons et al.(2013)]{parsons13} Parsons, S.~G., G{\"a}nsicke, B.~T., Marsh, T.~R., et al.\ 2013, \mnras, 429, 256 
\bibitem[Passy et al.(2012)]{passy12} Passy, J.-C., Mac Low, M.-M., \& De Marco, O.\ 2012, \apjl, 759, L30 
\bibitem[Perryman et al.(2001)]{perryman01} Perryman, M.~A.~C., de Boer, K.~S., Gilmore, G., et al.\ 2001, \aap, 369, 339 
\bibitem[Rowell \& Hambly(2011)]{rowell11} Rowell, N., \& Hambly, N.~C.\ 2011, \mnras, 417, 93 
\bibitem[Seoane et al.(2013)]{seoane13} Seoane, P.~A., et al.\ 2013, arXiv:1305.5720
\bibitem[Sigurdsson et al.(2003)]{sigurdsson03} Sigurdsson, S., Richer, H.~B., Hansen, B.~M., Stairs, I.~H., \& Thorsett, S.~E.\ 2003, Science, 301, 193 
\bibitem[Silvotti et al.(2007)]{silvotti07} Silvotti, R., Schuh, S., Janulis, R., et al.\ 2007, \nat, 449, 189 
\bibitem[Sim et al.(2012)]{sim12} Sim, S.~A., Fink, M., Kromer, M., et al.\ 2012, \mnras, 420, 3003 
\bibitem[Vennes et al.(2011)]{vennes11} Vennes, S., Thorstensen, J.~R., Kawka, A., et al.\ 2011, \apjl, 737, L16
\bibitem[Veras et al.(2013)]{veras13} Veras, D., Mustill, A.~J., Bonsor, A., \& Wyatt, M.~C.\ 2013, \mnras, 431, 1686 
\bibitem[Villaver \& Livio(2007)]{villaver07} Villaver, E., \& Livio, M.\ 2007, \apj, 661, 1192 
\bibitem[Webbink(1984)]{webbink84} Webbink, R.~F.\ 1984, \apj, 277, 355
\bibitem[Wolszczan \& Frail(1992)]{wolszczan92} Wolszczan, A., \& Frail, D.~A.\ 1992, \nat, 355, 145 
\bibitem[Wu \& Lithwick(2011)]{wu11} Wu, Y., \& Lithwick, Y.\ 2011, \apj, 735, 109 
\bibitem[Zucker et al.(2007)]{zucker07} Zucker, S., Mazeh, T., \& Alexander, T.\ 2007, \apj, 670, 1326 
\bibitem[Zuckerman et al.(2010)]{zuckerman10} Zuckerman, B., Melis, C., Klein, B., Koester, D., \& Jura, M.\ 2010, \apj, 722, 725 
\end{thebibliography}
}
\end{document}